# Direct observation of flat bands in near-magic-angle twisted bilayer CVD graphene

*Gianluigi Baiardi[1,2], Alex Boschi[1], Giulia Piccinini[1,2 †], Vaidotas Mišeikis[1], Lorenzo Cavicchi[3], Aaron Bostwick[4], Chris Jozwiak[4], Eli Rotenberg[4], Kenji Watanabe[5], Takashi Taniguchi[6], Marco Polini[7], Fabio Beltram[2], Antonio Rossi[1] *, Stiven Forti[1], Sergio Pezzini[8], Camilla Coletti[1] **

[1] Center for Nanotechnology Innovation @NEST, Istituto Italiano di Tecnologia, P.za San Silvestro 12, 56127 Pisa PI, Italy.

[2] NEST laboratory, Scuola Normale Superiore, P.za San Silvestro 12, 56127 Pisa PI, Italy.

[3] Scuola Normale Superiore, P.za dei Cavalieri 7, 56126 Pisa PI, Italy.

[4] Advanced Light Source, Lawrence Berkeley National Laboratory, 6 Cyclotron Rd, Berkeley, CA 94720, USA.

[5] Research Center for Electronic and Optical Materials, National Institute for Materials Science, 1-1 Namiki, Tsukuba, Ibaraki 305-0044, Japan.

[6] Research Center for Materials Nanoarchitectonics, National Institute for Materials Science, 1-1 Namiki, Tsukuba, Ibaraki 305-0044, Japan.

[7] Dipartimento di Fisica dell'Università di Pisa, Largo Bruno Pontecorvo 3, I-56127 Pisa, Italy.

[8] Istituto Nanoscienze - CNR, NEST-SNS, Piazza San Silvestro 12, Pisa, PI 56127, Italy.

## ABSTRACT

Advances in chemical vapor deposition (CVD) growth have driven graphene crystal quality to unprecedented levels, yet it is still unknown whether this route can realize the fragile flat-band and correlated states of the magic-angle (MA) twisted bilayer graphene (TBG). Here, we report on the experimental observation by room-temperature nano-angle-resolved photoemission spectroscopy (nano-ARPES) of flat bands in a TBG sample close to the MA, assembled via a grow-and-stack protocol based on low-pressure CVD of graphene on copper. Our study indicates electronic bands fully comparable to those measured in exfoliation-based samples and determines the size of the largest near-MA domain to be compatible with electronic transport experiments, motivating further experiments on flat band physics in CVD-graphene.

## MAIN TEXT

Two-dimensional materials provide a versatile platform for exploring and exploiting novel electronic,[1–3] optical,[4–7] and quantum[8–10] phenomena. When stacked with a relative twist, they form moiré superlattices that host emergent correlated states, flat bands, and exotic phases of matter.[11] Landmark discoveries, such as superconductivity[12] and correlated insulators[13,14] in magic-angle twisted bilayer graphene (MATBG), have been made exclusively with exfoliated flakes. Indeed, mechanical exfoliation represents the method of choice,[15–23] mostly thanks to the extremely low defect density of the individual layers and the use of clean, dry van der Waals assembly techniques.[24] However, this reliance on exfoliated samples poses intrinsic limitations in scalability, reproducibility,[25] and device integration.[26,27] Growth approaches to 2D crystals production, such as chemical vapor deposition (CVD),[28–31] offer an alternative that is getting in par in terms of material quality[32–36] and promises to achieve reliable protocols for the deployment of standardized devices.[37–39] Having established the capability of producing moiré TBG devices,[40] a central open question is whether the fragile flat bands at MA, highly sensitive to charge and twist-angle disorder, can be realized in a scalable platform such as CVD-grown graphene. Here we demonstrate via nano-ARPES, the observation of flat bands in twisted bilayer graphene from CVD, establishing a pathway toward scalable moiré materials. In addition, we study the relaxation of the near-MA configuration toward Bernal (AB) stacking across structural defects in the heterostructure, determining the size of the largest domain to be device-compatible for further studies of electronic transport in grown-and-stacked samples.

The key features of the fabricated sample and its geometry for the nano-ARPES experiment are summarized in Fig. 1. We selected two CVD-grown monolayer graphene single crystals, each a few hundred micrometers wide and sharing the same crystallographic orientation due to seeded growth on the same Cu grain.[41] These crystals were used as candidates for the grow-and-stack assembly[40] of MATBG (Fig. 1a). The final structure is schematically depicted in Fig. 1b, where the twist angle between the two graphene layers is increased for clarity, making the moiré unit cell (red hexagon) more visible. The pick-and-flip technique described in the Materials and Methods section of the Supporting Information (SI) allowed us to obtain an exposed TBG stack on top of a thin (~30 nm) hexagonal boron nitride (hBN) flake, ensuring atomic flatness (an optical image of the heterostructure is shown in Fig. S1a). The Raman mapping highlights the candidate MA areas (strain soliton contribution to the 2D peak higher than that of Bernal-stacked regions, Fig. S1c)[42] that were subjected to AFM tip-assisted brooming.[43] After cleaning the sample surface, we performed scanning tunneling microscopy (STM) measurements and observed a clear near-MA moiré pattern, as shown in Fig. 1c. The panel reports room-temperature STM topographic imaging of the moiré superlattice, along with a denoised portion obtained using a low-pass filter to enhance its visibility. Bright yellow regions corresponding to local AA stacking are connected by a triangular network of strain solitons (bright blue), similar to the hBN-unaligned MATBG shown in the work by Chen *et al*.[44] The moiré lattice parameter determined by the computation of the 2D autocorrelation function evaluates to 12.9 nm, corresponding to a twist angle of about 1.11°.

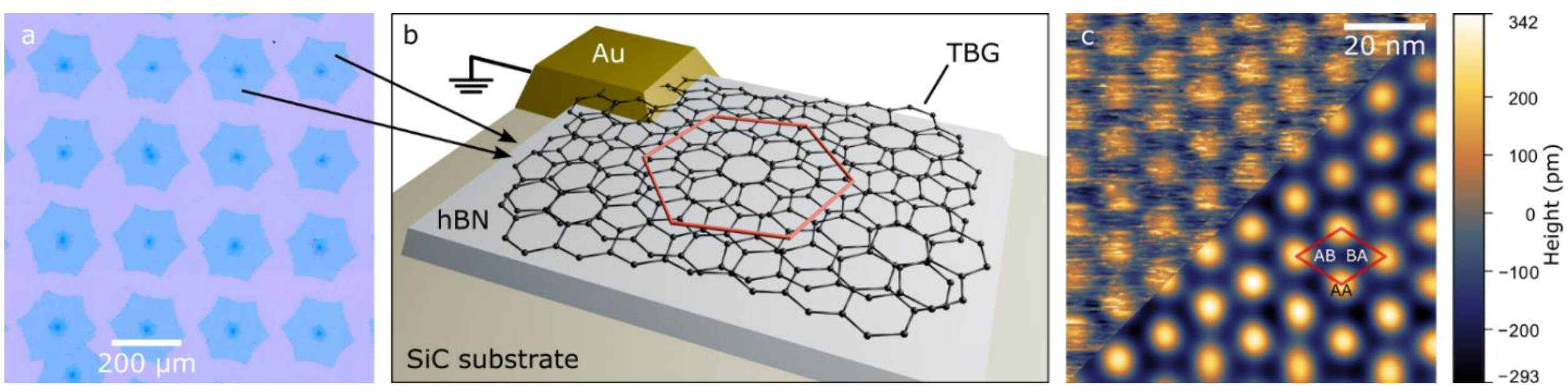

**Figure 1. a** Graphene crystals previously grown on the same Cu grain and transferred on a $Si/SiO_2$ substrate; precise control of the stacking angle is ensured by their mutual alignment. Two of them are highlighted as designed precursors for the fabrication of the near-MATBG device. **b** The architecture of the assembled device, featuring an exposed TBG structure (the twist angle is exaggerated for readability) on hBN. A red hexagon highlights the moiré superlattice unit cell. **c** Room-temperature STM imaging of the moiré superlattice, denoised with a low-pass filter in the bottom-right half. A red diamond identifies the primitive cell, containing the two Bernal-stacked regions (labeled in white) and spanning from one AA-stacked spot to another (labeled in black). The STM tip was biased at 50 mV, the current setpoint was 0.2 nA.

Once near-MA domain candidates were identified, we performed nano-ARPES at the ALS synchrotron directly on these regions. Full details of the measurements are provided in the Materials and Methods section. Fig. 2 reports three cuts across the valence-band photoemission intensity recorded from the largest domain under the ~1 μm$^2$ spot size of the light beam, together with three iso-energy surfaces extracted at different binding energies. Fig. 2a displays the ARPES-revealed electronic dispersion of the sample along the direction connecting the closest couple of single-layer Dirac cones, here labeled as $k_x$ and also referred to as the κmκ' direction in the reference frame of the moiré mini-Brillouin zone. In our spectra, the bands always appear as p-doped, with the shift from the expected neutrality point exceeding 100 meV where the local density of states is lowest, such as in 30°-TBG[45] reference regions (see Fig. S2a). This shift was witnessed also in an analogous sample deposited on Si/$SiO_2$ wafer and could, in principle, result from residual contamination introduced during growth, transfer, or stacking. However, in-house room- and low-temperature transport measurements on multiple devices fabricated with the same protocol all confirm the electrical neutrality of the graphene layers and rule out this scenario. In Fig. S2 we compare the carrier concentration in the present sample, as computed with nano-ARPES data, to the one in an exemplary monolayer graphene device, from transport measurements, demonstrating the neutrality of the material when it is not subjected to ionizing radiation. Considering the reported data and analogous ARPES results in other experiments,[46,47] we attribute the effect to synchrotron-radiation-induced photodoping. We point out that the details of this photodoping effect are not fully clarified; dedicated studies beyond the scope of this work would be required to investigate its dependence on photon flux and possible transient behavior. Regarding the latter, both the sample investigated here, and the analogous sample deposited on Si/$SiO_2$ exhibited a consistent level of doping throughout the measurement session, with no obvious evolution of the observed band structure. While this observation is compatible with a rapid establishment of the photodoped state under synchrotron illumination, the available data does not allow us to draw firm conclusions on the underlying dynamics. As a consequence of photodoping, the electronic states near the charge-neutrality point were not accessible in our measurements. Nonetheless, by taking a high-symmetry cut slightly off the κmκ' axis (i.e. along a parallel line, as in Fig. 2a) it is possible to observe the onset of the flat bands near the Fermi level: due to the extremely high density of states, the photo-induced carriers reside within a narrow energy window, resulting in partially hole-populated flat bands that are clearly visible (albeit off the K points of graphene). A comparison between the bands observed in our sample along this direction and the standard electronic structure of MATBG computed via the Bistritzer-MacDonald model[48] is reported in Fig. S3, showing overall similarity. The morphology of the signal distribution throughout the whole of Fig. 2 is fully comparable to that previously reported in other works for devices close to the MA fabricated from exfoliated graphene.[49,50] Specifically, the orthogonal cuts along the $k_y$ direction in Fig. 2b-c (mγm' direction) clearly display, from lower to higher binding energy, a flattening of the low-energy band close to the Fermi level and a gap opening resulting from band hybridization at low energies. Moreover, we note that the ARPES intensity from the inner bands is stronger than for the outer ones (Fig. 2b), consistently with the spectral weight shift recently reported by Li *et al.*[51] for MATBG, adding to the experimental evidence obtained so far in favor of the near-MA configuration. To place our results in the context of previous nano/microARPES studies, we performed a quantitative analysis of representative energy and momentum distributions curves (EDCs and MDCs) extracted from Fig. 2b. The resulting flat-band linewidth and dispersion, gap-related energy scale, and MDC linewidths are summarized in Table S1, while the fitting procedures are reported in Fig. S4. We point out that the flat band width is not univocally defined in the literature (see Table S1), which also includes substantial differences in experimental conditions among the available studies. Nevertheless, the extracted spectral metrics are broadly consistent with those reported for exfoliated MATBG samples. Isoenergetic intensity distributions are provided in Fig. 2d-f for three energy values, illustrating the Fermi surface and the development of an articulated structure derived from the interaction of the Dirac cones close to the MA. Fig. 2e-f exhibit an omega-like shape, modulated by the photoemission matrix element, fully compatible with previous studies on near-MA samples.[49] Overall, the electronic structure measured in reciprocal space indicates that CVD grown-and-stacked graphene can host delicate moiré flat-band configurations, underscoring the high quality achievable with current growth and transfer methods.

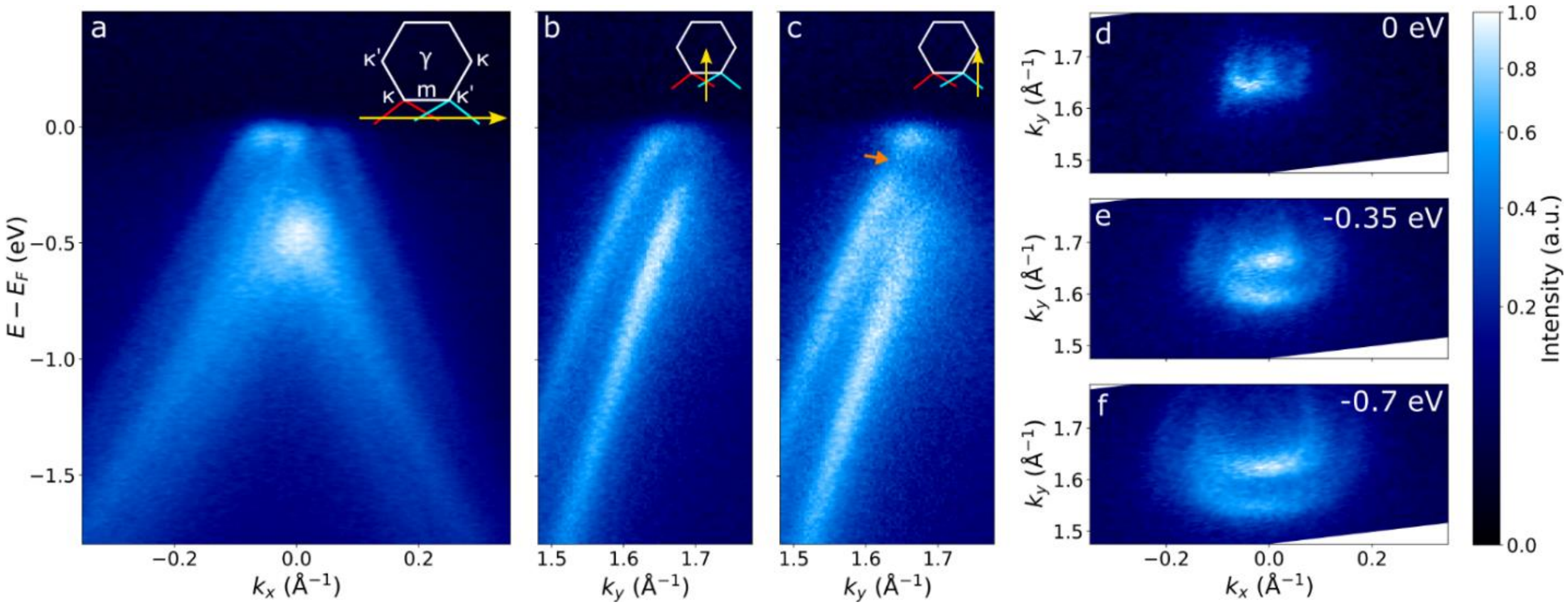

**Figure 2.** Cuts across the ARPES intensity recorded inside a near-MATBG domain with a photon energy of 27 eV and a pass energy of 30 eV, at room temperature, with 1 μm$^2$ spot size. Coordinates are referred to a single graphene layer Brillouin zone. The insets show a TBG mini-Brillouin zone (white), where the red and blue segmented lines belong to the original first Brillouin zones of the twisted graphene layers. The first inset is enlarged for readability; the rest are on scale with the axes. Notice the non-linear color scale, mapped to the square root of normalized intensity values to enhance low-intensity features. **a** Cut along the $k_x$ (κmκ') direction, slightly off the intersection with the Dirac points of the individual graphene layers. The low-energy hybridization of the original cones allows us to observe the onset of the flat bands, while at higher energies a nontrivial structure featuring multiple bands is present. **b, c** Cuts along the $k_y$ (mγm') direction, at $k_x$ values indicated by the yellow arrow in the inset. The flattening of the low-energy band is evident in panel b, while a hybridization-driven gap opens about –0.2 eV in panel c (orange arrow). **d-f** Isoenergetic cuts at the Fermi level (0 eV), at –0.35 eV, and at –0.7 eV, showing the complex photoemission intensity distribution arising from the hybridization of the Dirac cones, analogous to already reported data in the literature for MATBG devices built out of exfoliated graphene flakes.

Having established the presence of flat bands, we next examine their spatial distribution, as well as the evolution due to relaxation of the twist angle, leveraging the 1-μm$^2$ resolution of nano-ARPES. To this end, Fig. 3 displays the analysis of a single hyperspectral dataset. In Fig. 3a, the signal at each location is obtained by integrating the ARPES intensity over a shallow window around the origin of the $k_x$ and $E$ axes, covering 0.15 Å$^{-1}$ in momentum and 40 meV in energy (its perimeter is shown as a yellow rectangle in panels b-d). The binning size is enough to capture the full contribution of the flat band region and distinguish the lower signal provided by the steeper dispersion of Bernal-stacked areas. As a result, locations endowed with near-MA dispersion stand out from the rest and allow initially estimating in 15×10 $\mu m^2$ the extension of the near-MA domain under observation. To further strengthen the identification of electronically different domains in the sample, we implemented an unsupervised machine-learning analysis based on Non-negative Matrix Factorization (NMF). NMF decomposes the hyperspectral dataset into a small number of non-negative spectral components and their corresponding spatial distributions. Compared to Principal Component Analysis (PCA), NMF is particularly suitable for photoemission data because both the spectral intensity and the extracted components are constrained to be positive, resulting in a more physically interpretable decomposition in terms of distinct electronic contributions. NMF identifies four domains, represented in Fig. 3a via solid contour lines, that match the three representative dispersions in Fig. 3b-d, plus a fourth region with scarce signal (due to the physical edge of the sample, bottom-right, and to the lack of the TBG structure, top-right). Details on the procedure are reported in the SI. The orange dotted square in Fig. 3a highlights the AFM-broomed area, which displays substantially less noise than the surroundings in the ARPES signal acquired. The same panel includes as a red dashed line the position of a wrinkle in the hBN flake which likely occurred during fabrication. The asperity induced by this defect appears to cause the near-MA configuration to fully relax to Bernal stacking at that location and beyond. As is visible in Fig. S6, across the wrinkle the overall bilayer orientation on the substrate experiences discontinuities that locally misorient both layers of graphene, leading to their rearrangement in the most favorable configuration regardless of the angle imposed during the assembly. This wrinkles-induced relaxation highlights a residual degree of uncertainty in the outcome of current state-of-the-art dry assembly techniques, that can be mitigated only up to a certain extent.[15,52] To increase the MATBG yield and domain size, one could devise new stacking schemes that reduce as much as possible the human factor[53], as well as the mechanical stress on the crystals (see for instance the low-pressure technique recently introduced to preserve rhombohedral stacking[54]).

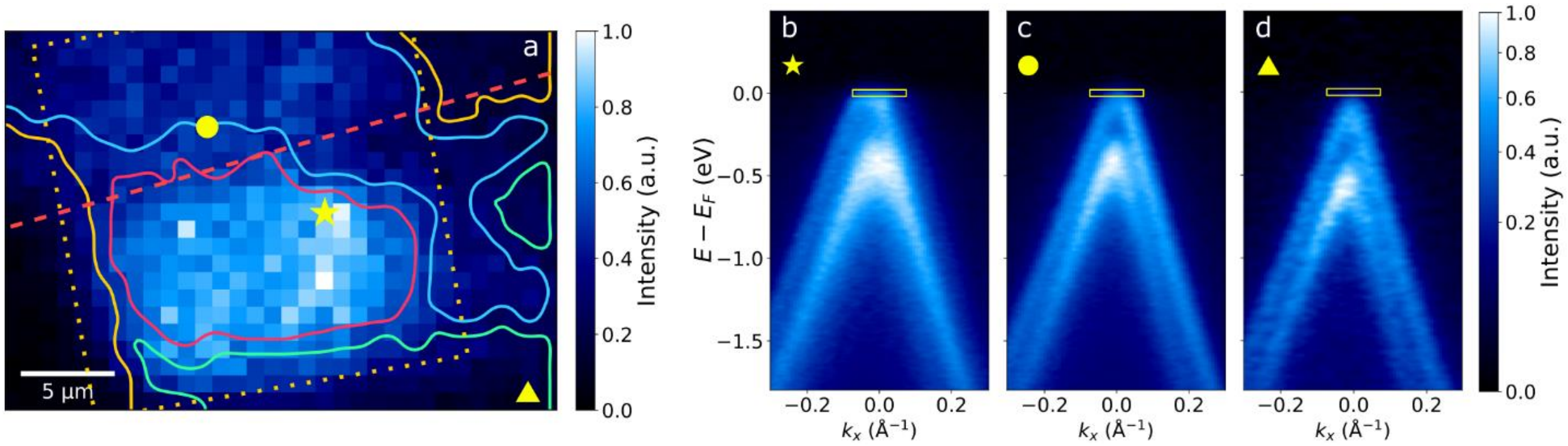

**Figure 3. a** Real-space imaging of the largest near-MA domain in the sample, highlighted by integrating the room-temperature ARPES intensity on a 0.15 Å$^{-1}$-wide and 40 meV-deep window about $k_x$=0 and the Fermi level, displayed in yellow in panels b-d. The spatial distribution of the spectral components identified by NMF applied to the nano-ARPES hyperspectral dataset is displayed with solid-line contours at 80% assignment probability obtained via a Gaussian Mixture Model. The analysis separates a near-MA region (red contour), a Bernal-stacked region within the AFM-broomed area (cyan contour), and a Bernal-stacked region outside the broomed area (green contour). The orange dotted square marks the AFM-broomed area, while the red dashed line indicates the position of a wrinkle in the underlying hBN flake **b-d** Representative κмκ'-parallel ARPES cuts acquired at the positions marked in panel (a), corresponding to the near-MA region (b), the broomed Bernal region (c), and the nonbroomed Bernal region (d). The latter exhibits a shifted apparent Fermi level, consistent with a stronger influence of residual surface contamination. All cuts are acquired off the K points of both graphene layers as in Fig. 2. The color scale is identical to that of Fig. 2. Data were acquired with a photon energy of 95 eV and a pass energy of 50 eV.

To obtain a more quantitative estimation of the length scale over which the MA configuration of this large domain relaxes, an approach based on fitting of position-resolved EDCs was adopted (Fig. 4). This method allows pinpointing the evolution of the low-energy band by building a more robust indicator of its shape than the simple integrated intensity as in Fig. 3. Initially, binned EDCs have been extracted across the near-MA domain along its short side (see Fig. S6a for the exact direction of the cut). The binning in $k_x$ was set again to 0.15 Å$^{-1}$ around the center of the $k_x$ axis (Fig. S6b), to concentrate the spectral weight of the flat band into a narrow energy range, while diluting contributions from Bernal-like and higher-energy bands. In this way, flatter dispersions manifest as sharper peaks in the binned EDCs (Fig. S6c). Fig. 4a displays the first series of position-resolved EDCs, extracted along the white arrow in Fig. S6a, with the color-mapped ARPES intensity separately normalized within each EDC. The onset of the flat band is clearly visible in the raw data in the central region of the plot (yellow arrow), owing to their strong intensity. However, to take into account Fermi distribution effects, the next step is to fit each EDC as exemplified in Fig. 4b-c for two opposite cases, corresponding to a Bernal-relaxed location out of the near-MA domain and a site inside the domain, respectively. The procedure includes a prior Shirley background subtraction and the fitting of two unconstrained components, built as a product of a Lorentzian profile and the Fermi distribution, subsequently convolved with a Gaussian distribution to include instrumental uncertainty (see SI for details). The Lorentzian profile center position of the weaker component (in blue in Fig. 4b,c) for each EDC is displayed in Fig. 4d and constitutes the desired indicator of the electronic configuration shape across the near-MA domain: in the middle of the domain, the flat band yields a photoemission intensity concentrated just below the Fermi level throughout the $k_x$-binning range and thus assigned by the fitting algorithm to a component centered at the Fermi level; conversely, where the low-energy band is more dispersive, the center position of the associated component is shifted in energy to more negative values, eventually saturating for Bernal-stacked areas where the dispersion is the steepest. To improve consistency, Fig. 4d aggregates the results of five parallel series of EDCs (the remaining four are reported in Fig. S7, their directions being the green-shaded arrows in Fig. S6a) and thus displays the average position of the low-energy band, with $\pm\sigma$ as the shaded area. We note that the method performs better inside the less noisy and doped AFM-broomed region (i.e. up to ~17 μm), while out of it the curve doesn't seem to reach a stable value. Considering the apparent bell-like decay of the magic-angle configuration, its relaxation length can be described as half FWHM, equal to 4.1±1.1 μm. With this analysis being done on the shorter side of the domain, this result marks an essential achievement in view of fabricating MATBG devices starting from graphene synthesized by a scalable technique such as CVD. By performing the same procedure along the other axis of the domain, it is possible to give a new estimation of the domain size as 84±41 $\mu m^2$. The high uncertainty is due both to the measurement resolution, relatively low with respect to the size of the near-MA domain, and to limitations of the method itself, which possibly trades fine details of the electronic dispersion for an improved signal. Nevertheless, this surface extension represents a promising scale for studying the homogeneity of the electronic properties[55] in electrical transport experiments.

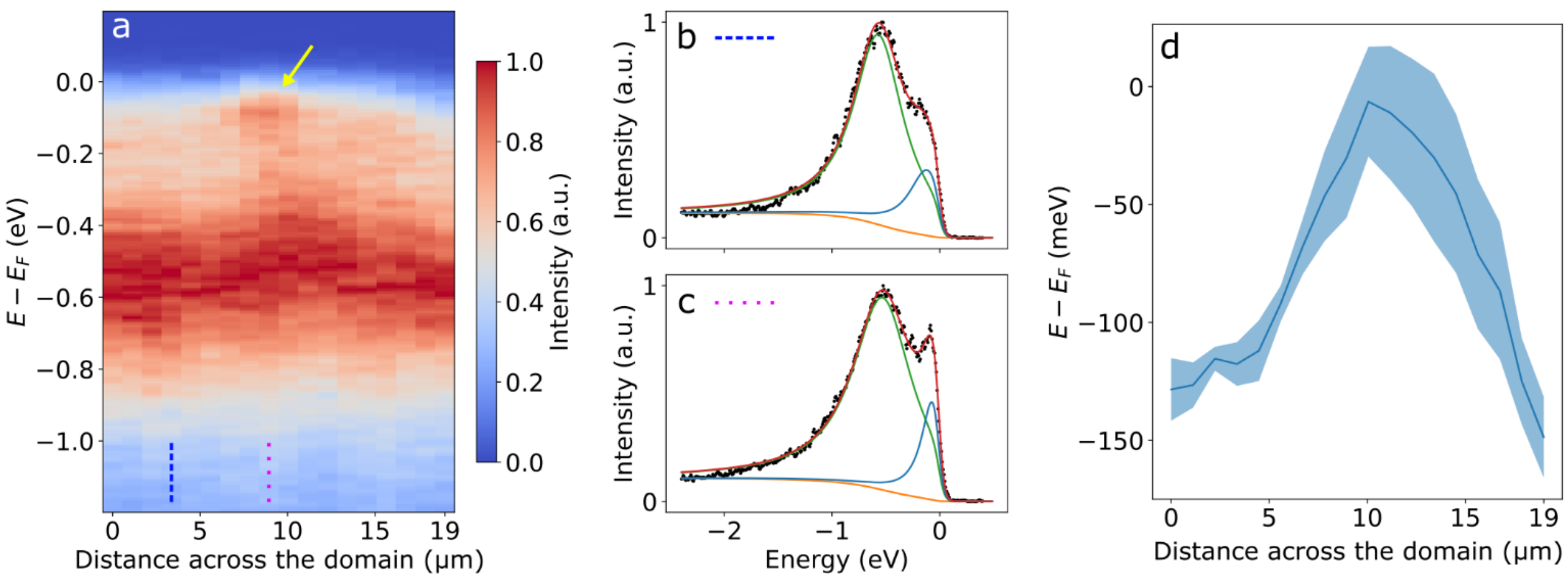

**Figure 4.** Estimation of the magic-angle configuration relaxation length. **a** $k_x$-binned EDCs extracted across the near-MA domain (along the white arrow in Fig. S6a). The largest contribution extending about –0.5 eV in each EDC is due to the integrated signal of the remote band. A weaker signal, due to the low-energy band, is responsible for the still relevant intensity close to the Fermi level, in few cases standing out as a prominent peak (marked by the yellow arrow). **b, c** Two examples of binned EDCs from the positions marked by the blue dashed and pink dotted lines in panel a. The fitting algorithm first subtracts a Shirley background, then fits two components built as products of a Lorentzian lineshape and the Fermi distribution, subsequently convolved with a Gaussian profile. **d** The optimized center position of the low-energy fitted component (blue lineshape in panels b and c) against the real-space axis of extraction of the binned EDCs. A value close to the Fermi level is the fingerprint of the flat band, while larger negative values correspond to more and more dispersing bands.

The present study reports on the realization of the fragile near-MA configuration in exposed TBG by stacking iso-oriented graphene crystals grown via CVD. In addition to the observation of a moiré pattern by STM and an ARPES signal fully compatible with those achieved with exfoliated graphene flakes in the literature, a spatial analysis is carried out to further characterize the extension of the largest near-MA domain obtained of about 84 μm$^2$. The observation and quantitative characterization of a clean and structurally coherent near-MA domain underline the maturity reached by our fabrication protocol and provides compelling evidence that scalable CVD graphene is capable of hosting the fragile flat-band electronic structure previously accessible only in exfoliated systems. Addressing the fabrication of transport devices suitable for investigating correlated electronic phases will, however, require overcoming additional challenges associated with the integration of contacts and gates. Considering our results and recent advancements in the fabrication of CVD-based devices[32,34,35] and TBG,[36,40] this objective appears to be within reach. This work thus marks a significant step toward enabling future investigations of correlated quantum states in large-area, synthetically grown graphene.

## SUPPORTING INFORMATION

The following files are available free of charge.

- Materials and Methods section; Unsupervised Machine Learning-enabled procedure to automatically detect domains in the sample; Detailed procedure for evaluating the near-MA configuration relaxation length; Continuum-model calculation, band unfolding, and ARPES spectral weight in twisted bilayer graphene; Supplementary table S1; Supplementary figures S1-S7 (PDF).

## AUTHOR INFORMATION

### Corresponding Author

* Email: antonio.rossi@iit.it * Email: camilla.coletti@iit.it.

### Present Addresses

† ICFO - Institut de Ciències Fotòniques, The Barcelona Institute of Science and Technology, Castelldefels (Barcelona), Spain.

### Author Contributions

G.P. and S.P. fabricated the sample employing the material provided by V.M., K.W., and T.T.; A. Boschi performed the Raman analysis and AFM brooming; S.F. conducted the STM imaging; A.R. and S.F. performed the ARPES measurement with the support of A. Bostwick, C.Z., and E.R.; L.C. carried out the MATBG electronic structure computation; G.B. analyzed all data and wrote the first draft of the manuscript; A.R., S.P., and C.C. coordinated the experimental work. All authors contributed to the manuscript.

### Notes

The authors declare no competing financial interests.

Data presented in this article, along with non-proprietary software employed to process and plot it, are available at the following link: https://doi.org/10.48557/MY2BCP.

## ACKNOWLEDGMENTS

We acknowledge support from Project "National Quantum Science And Technology Institute", PE000023 funded by the European Union - NextGenerationEU PNRR MUR - M4C2 - Investimento 1.3 - Avviso Creazione di "Partenariati estesi alle università, ai centri di ricerca, alle aziende per il finanziamento di progetti di ricerca di base" CUP J53C22003200005.

A. Boschi acknowledges funding from the European Union through the GraPh-X project (Grant number 101070482).

K.W. and T.T. acknowledge support from the JSPS KAKENHI (Grant Numbers 21H05233 and 23H02052), the CREST (JPMJCR24A5), JST and World Premier International Research Center Initiative (WPI), MEXT, Japan.

This research used resources of the Advanced Light Source, which is a DOE Office of Science User Facility under contract no. DE-AC02-05CH11231.

Artificial Intelligence (AI) tools and, specifically, Large Language Models (LLMs) were employed to develop the newly written code necessary to perform the data analysis.

**SUPPORTING INFORMATION FOR**

# Direct observation of flat bands in near-magic-angle twisted bilayer CVD graphene

*Gianluigi Baiardi[1,2], Alex Boschi[1], Giulia Piccinini[1,2 †], Vaidotas Mišeikis[1], Lorenzo Cavicchi[3], Aaron Bostwick[4], Chris Jozwiak[4], Eli Rotenberg[4], Kenji Watanabe[5], Takashi Taniguchi[6], Marco Polini[7], Fabio Beltram[8], Antonio Rossi[1] *, Stiven Forti[1], Sergio Pezzini[8], Camilla Coletti[1] **

[1] Center for Nanotechnology Innovation @NEST, Istituto Italiano di Tecnologia, P.za San Silvestro 12, 56127 Pisa PI, Italy.

[2] NEST laboratory, Scuola Normale Superiore, P.za San Silvestro 12, 56127 Pisa PI, Italy.

[3] Scuola Normale Superiore, P.za dei Cavalieri 7, 56126 Pisa PI, Italy.

[4] Advanced Light Source, Lawrence Berkeley National Laboratory, 6 Cyclotron Rd, Berkeley, CA 94720, USA.

[5] Research Center for Electronic and Optical Materials, National Institute for Materials Science, 1-1 Namiki, Tsukuba, Ibaraki 305-0044, Japan.

[6] Research Center for Materials Nanoarchitectonics, National Institute for Materials Science, 1-1 Namiki, Tsukuba, Ibaraki 305-0044, Japan.

[7] Physics department, Università di Pisa, Largo Bruno Pontecorvo 3, 56127 Pisa PI, Italy.

[8] NEST laboratory, Istituto Nanoscienze-CNR and Scuola Normale Superiore, P.za San Silvestro 12, 56127 Pisa PI, Italy.

# Materials and methods

Graphene single crystals were epitaxially grown on electropolished and annealed copper foil via low-pressure CVD[1] (in a BM® reactor by Aixtron), and subsequently semi-dry-transferred with the aid of a polymeric stack (prepared with poly(propylene carbonate), PPC, and polymethyl methacrylate, PMMA)[2] on a silicon wafer substrate (with 285-nm thermal oxide) for characterization and selection.

Hexagonal boron nitride (hBN) was exfoliated via a parent/child-tape procedure[3] (Magic tape® by 3M) and released on analogous Si/$SiO_2$ chips after a short annealing on a hotplate in air (90°C for 1 minute) to increase the release yield.

The selection of graphene crystals was performed by optical microscopy and Raman spectroscopy (employing a 532-nm green laser and an InVia® microscope by Renishaw), both to ensure their cleanness from PMMA residuals and to avoid defective and strained crystals. Two distinct nearby crystals, grown on the same Cu grain and thus sharing the same crystalline orientation, were chosen as original monolayers for the assembly of MATBG. A single hBN flake featuring a flat, uncracked, glue-free area of several tens of micrometers in lateral size was selected by optical microscopy and atomic force microscopy (AFM, with a Dimension Icon® setup by Bruker).

The exposed TBG sample was fabricated via a modified version[4] of the pick-and-flip method originally developed by Wong *et al.*,[5] which incorporates the dry-transfer technique enabled by a polycarbonate/polydimethylsiloxane (PC/PDMS) stamp placed on a glass slide.[6] In this modified procedure, a polyvinyl alcohol (PVA) membrane is deposited on top of the standard stamp and is used to sequentially pick all layers up in bottom-up order, i.e. hBN, bottom graphene, and top graphene; the heterostructure is subsequently brought into contact with a second PC/PDMS stamp (PC by Sigma Aldrich, PDMS Sylgard 184), so that dissolving the PVA film in a water drop allows to flip the sample, which is thermally released on a clean substrate following Purdie *et al*.[6] All involved steps were performed with a custom-built transfer setup, and a twist angle of 1.2° was targeted to compensate for the tendency of TBG to relax towards the Bernal stacking. The final substrate was a silicon carbide chip, chosen both for the atomic flatness of the terraces exposed on the surface after etching, and as part of an attempt to mitigate the photodoping effect previously witnessed at the synchrotron on samples deposited on Si/$SiO_2$ wafers.

The exposed TBG was then characterized by all aforementioned techniques, identifying the most promising areas as those featuring a contribution of the strain soliton region (AB-to-BA stacking domain boundary) to the Raman 2D peak at least as high as that of the Bernal regions (see Fig. S1), as suggested by Barbosa *et al.*[7] Cr/Au contacts were evaporated on TBG for electrical grounding. The regions of interest were eventually broomed with prolonged AFM scans in contact mode (with an afm+® setup by Anasys Instruments) to further clean them from PC and lithography resist residuals and improve the signal-to-noise ratio during both scanning tunneling microscopy (STM) and nano-ARPES investigations.

STM imaging was accomplished in-house (LT-STM by Scienta Omicron) at room temperature to confirm the near-magic-angle configuration in the regions of interest by direct observation of the moiré pattern. Nano-ARPES measurements were carried out at the MAESTRO beamline of the Advanced Light Source (ALS) at Lawrence Berkeley National Laboratory (LBNL), California, US. Long-exposure snapshots of the TBG electronic dispersion along $k_x$, reciprocal-space maps of the full 2D dispersion, and hyperspectral maps with up to 1-μm$^2$ resolution were all acquired at energies of 27 and 95 eV, respectively with a pass energy of 30 and 50 eV, enabling the spectroscopic characterization of the presented device and the subsequent discussion. For both STM and nano-ARPES experiments a vacuum annealing was carried out at 150°C overnight.

AFM and STM images were processed with Gwyddion 2.68 standard modules.[8] Raman data were elaborated with Renishaw's WiRE proprietary software and, with respect to the more in-depth analysis reported in the SI, by custom Python codes, available on the linked repository (see Data Availability). ARPES data were analyzed via BEAMLINE 7.0.2 Igor Pro routines, other custom Igor Pro routines, and custom Python codes.

With respect to reproducibility, we wish to note that both the sample discussed here and its counterpart on $Si/SiO_2$, mentioned in the main text, exhibited near-MA hybridization of the Dirac cones, albeit with different domain sizes. An identical data-processing procedure yields an estimated largest near-MA domain of approximately 10×5 μm$^2$ for the $Si/SiO_2$-supported sample. Among the three samples that were successfully transferred, assembled, and subsequently characterized by nARPES, two exhibited near-MA electronic structure. Furthermore, the largest electronically defined near-MA domain increased from approximately 10×5 μm$^2$ in the $Si/SiO_2$-supported sample to 84±41 μm$^2$ in the sample discussed here.

# Unsupervised Machine Learning-enabled procedure to automatically detect domains in the sample

Non-negative Matrix Factorization (NMF) allows to factorize a positively-valued matrix N×M, where the N rows represent the items in a collection, the M columns represent the features of such items, and the matrix elements represent the amount of such features in every item, into a pair of positively-valued matrices N×A and A×M, where the arbitrarily long dimension vanishing with the product represents a number of fundamental patterns of features present in the collection of items, able to reconstruct the original dataset with the least possible error (the algorithm leading to the factorization is indeed a compression algorithm which minimizes the distance with the original N×M matrix). The non-negativity of such matrices opens to the interpretation of the rows of the A×M matrix as the description of such patterns in terms of the features, and of the rows of the N×A matrix as the relative weight of each pattern within each item in the collection.

We applied such framework to identify domains in our sample, via ARPES intensity and within the field of view given by Fig. 3a, by treating the x-y position as the items and the k-E dispersed signal as the features. The N×M matrix was obtained by cropping the wavevector and energy axes to the most relevant portions (60 and 145 values respectively) and flattening the 4D dataset into a 704×8700 two-dimensional matrix. The optimal number A of patterns, here electronic dispersion signatures, can be fine-tuned iteratively but an educated guess is possible based on direct inspection of the dataset. We found that 3 signatures yielded the optimal result. The assignment of items (x-y positions) to clusters, in order to identify and visualize the domains, was done by a Gaussian-Mixture Model, preferred to a simple k-means clustering as it enables a probabilistic assignment that keeps track of the smoothness of the transition between clusters. Again, the optimal number of clusters can be found iteratively, but asking for 4 clusters yielded the most sensible result, where each of the first three clusters directly corresponds to a specific electronic signature, and the fourth contains no relevant content of any signature, due to degraded ARPES signal beyond the edge of the sample (bottom-left) or the lack of the TBG structure (top-right).

# Detailed procedure for evaluating the near-MA configuration relaxation length via EDCs

1. Define the direction of interest while imaging the domain via spectral data (Fig. S5a).
2. Extract the k- and E-resolved photoemission data corresponding to the relevant positions. Remember to correct the position of the Fermi level, if not calibrated yet, for the fitting algorithm to yield the correct result.
3. Compute binned EDCs by summing the photoemission intensity over a suitable kx interval.
4. For each EDC fit a Shirley background and a number of components equal to the number of bands visible in the photoemission data. To be physically consistent, the components should be the product of a Lorentzian profile and a Fermi distribution function, subsequently convolved with a same-width Gaussian distribution to account for instrumental uncertainty.
5. If possible, to improve the accuracy of the method, repeat the extraction and fitting along parallel axes across the domain, then average the results position-wise.
6. Display the position (in energy) of the Lorentzian profile vs the real-space coordinate marked by the extraction axis.

# Continuum-model calculation, band unfolding, and ARPES spectral weight in twisted bilayer graphene

Twisted bilayer graphene (TBG) was described within the standard single-valley continuum model in the layer-sublattice basis $\{|1A\rangle, |1B\rangle, |2A\rangle, |2B\rangle\}$. The single-particle Hamiltonian was written as[9]

$$\hat{H} = \begin{pmatrix} \hat{H}^{(t)} & \hat{U} \\ \hat{U}^{\dagger} & \hat{H}^{(b)} \end{pmatrix}$$

where $\hat{H}^{(\ell=t,b)}$ is the rotated Dirac Hamiltonian of layer ℓ and U is the moiré interlayer tunneling operator. The intralayer blocks were taken in the low-energy $k \cdot p$ form (± sign refers to valley)

$$\hat{H}^{(\ell)} = v_D[R_\ell(\theta/2)(\hat{\boldsymbol{p}} \mp \hbar \boldsymbol{K}_\ell)] \cdot \left(\pm\sigma_x, -\sigma_y\right),$$

and the interlayer tunneling was expanded in the three leading moiré harmonics,

$$\hat{U} = T_0 + T_1\, exp(i\boldsymbol{G}_1 \cdot \hat{\boldsymbol{r}}) + T_2\, exp(i\boldsymbol{G}_2 \cdot \hat{\boldsymbol{r}}),$$

with $T_j$ parameterized by the intrasublattice and intersublattice hopping amplitudes $u_0$ and $u_1$.

The eigenstates were obtained by diagonalizing $\hat{H}$ in a plane-wave basis. For crystal momentum $\boldsymbol{\kappa}$ in the first moiré Brillouin zone (mBZ), the Bloch eigenstates were expanded as

$$\langle \boldsymbol{r}|\boldsymbol{\kappa}, \lambda\rangle = \frac{1}{\sqrt{S}} \sum_{\boldsymbol{G}} u_G(\boldsymbol{\kappa}, \lambda)\, exp[i(\boldsymbol{\kappa} + \boldsymbol{G}) \cdot \boldsymbol{r}],$$

where $\boldsymbol{G}$ belongs to the moiré reciprocal lattice and $u_{\boldsymbol{G}}(\boldsymbol{\kappa}, \lambda)$ is a four-component spinor in layer/sublattice space.

To unfold the moiré bands into the primitive Brillouin zone (pBZ) of graphene, primitive-cell momenta were constructed from a given moiré momentum $\boldsymbol{\kappa}$ according to[10]

$$\boldsymbol{k}_i = \boldsymbol{\kappa} + \boldsymbol{G}_i,$$

where $\boldsymbol{G}_i$ runs over the moiré reciprocal lattice vectors that map the state back into the pBZ. The unfolded spectral weight was defined as the projection of the supercell eigenstate onto the primitive-cell Bloch subspace,

$$P_{\boldsymbol{\kappa}\lambda}(\boldsymbol{k}_i) = \sum_\nu |\langle \boldsymbol{k}_i, \nu|\boldsymbol{\kappa}, \lambda\rangle|^2 .$$

Using the plane-wave decomposition above, this weight can be evaluated directly from the moiré eigenvectors as

$$P_{\boldsymbol{\kappa}\lambda}(\boldsymbol{k}_i) = \sum_{\boldsymbol{g} \in pRL} \left|u_{\boldsymbol{g}+\boldsymbol{G}_i}(\boldsymbol{\kappa}, \lambda)\right|^2 ,$$

where the norm includes the sum over the internal layer and sublattice components.

The unfolded spectral function was then computed as[10,11]

$$A(\boldsymbol{k}_i, E) = \sum_{\lambda} P_{\boldsymbol{\kappa}\lambda}(\boldsymbol{k}_i)\delta(E - \varepsilon_{\boldsymbol{\kappa}\lambda}).$$

In the numerical implementation, the Dirac delta was replaced by a finite broadening. A Lorentzian form was used,

$$\delta(E - \varepsilon_{\boldsymbol{\kappa}\lambda}) \rightarrow \frac{\eta}{\pi[(E - \varepsilon_{\boldsymbol{\kappa}\lambda})^2 + \eta^2]} ,$$

with $\eta$ the chosen energy-resolution parameter.

Finally, to model ARPES spectra, a free-electron final-state approximation was adopted, and a layer interference factor was included explicitly. The resulting intensity was written as[12]

$$A(\boldsymbol{k}_i, E) = \sum_{\lambda} \sum_{\boldsymbol{g} \in pRL} \left| \sum_{\ell=t,b} \sum_{\alpha=A,B} u_{\alpha\ell,\boldsymbol{g}+\boldsymbol{G}_i}(\boldsymbol{\kappa}, \lambda)\, e^{ik_z d_\ell} \right|^2 \delta(E - \varepsilon_{\boldsymbol{\kappa}\lambda}).$$

Here $d_\ell$ is the vertical position of layer ℓ, taken symmetrically as $d_t, b = \pm d/2$ with $d \approx 0.34\ nm$. The out-of-plane photoelectron momentum was evaluated as[12]

$$k_z = \sqrt{\frac{2m}{\hbar^2}\left[\hbar\omega_{ph} - W + \varepsilon_{\boldsymbol{\kappa}\lambda}\right] - |\boldsymbol{k}_i|^2},$$

where $\hbar\omega_{ph}$ is the photon energy, $W$ is the work function, and $\boldsymbol{k}_i$ is the unfolded in-plane crystal momentum. This form captures the attenuation and phase interference between photoelectrons emitted from the two graphene layers.

# Supplementary table

| | (near-)MA domain size (μm$^2$) | EDC linewidth (meV) | Flat Band dispersion** (meV) | Moiré gap (meV) | MDC linewidths (Å$^{-1}$) | Background-to-flat-band ratio |
|---|---|---|---|---|---|---|
| Lisi et al., 2021 | ~50 * | | 30±15 | | | |
| Jiang et al., 2023 | ~50 * | 70 | | 140 | | |
| Li et al., 2024 | ~4 * | | 11±10 | | | |
| Chen et al., 2024 | | | | 150±15 | | |
| *This work* | 84±41 | ~118 | ~24 | ~155 | 0.008 – 0.080 | 0.877 |

* Very rough estimate of the domain size based on available spectral mapping images.

** Obtained by tracking the position of the maximum intensity across the k-space, not by estimating the linewidth of the low-energy band.

**Table S1** – Comparison between quantitative features of the electronic dispersion characterized in this work and published values for analogous near-MA and MA TBG systems.
We extracted and fitted energy dispersion curves (EDCs) at the m point and momentum dispersion curves (MDCs) at -0.5 eV from Fig. 2b. Details on the fitting are reported in Fig. S4. Being the moiré bands folding very sensitive to the twist angle, we observe that the size and depth of the gaps can be related to the average twist within the probed area, while the sharpness of the feature relates to the temperature and the homogeneity in the twist. As such, our domain shows a little higher disorder compared to analogous samples crafted with exfoliated flakes. Conversely, from the analysis of MDCs at -0.5 eV we report sharp bands, with FWHM values reaching 0.008 Å$^{-1}$. This could be explained by a more homogeneous overall orientation of the whole bilayer within the spot size of nARPES, leading to sharper dispersive bands as the Dirac cones do not shift in the reciprocal space.

# Supplementary figures

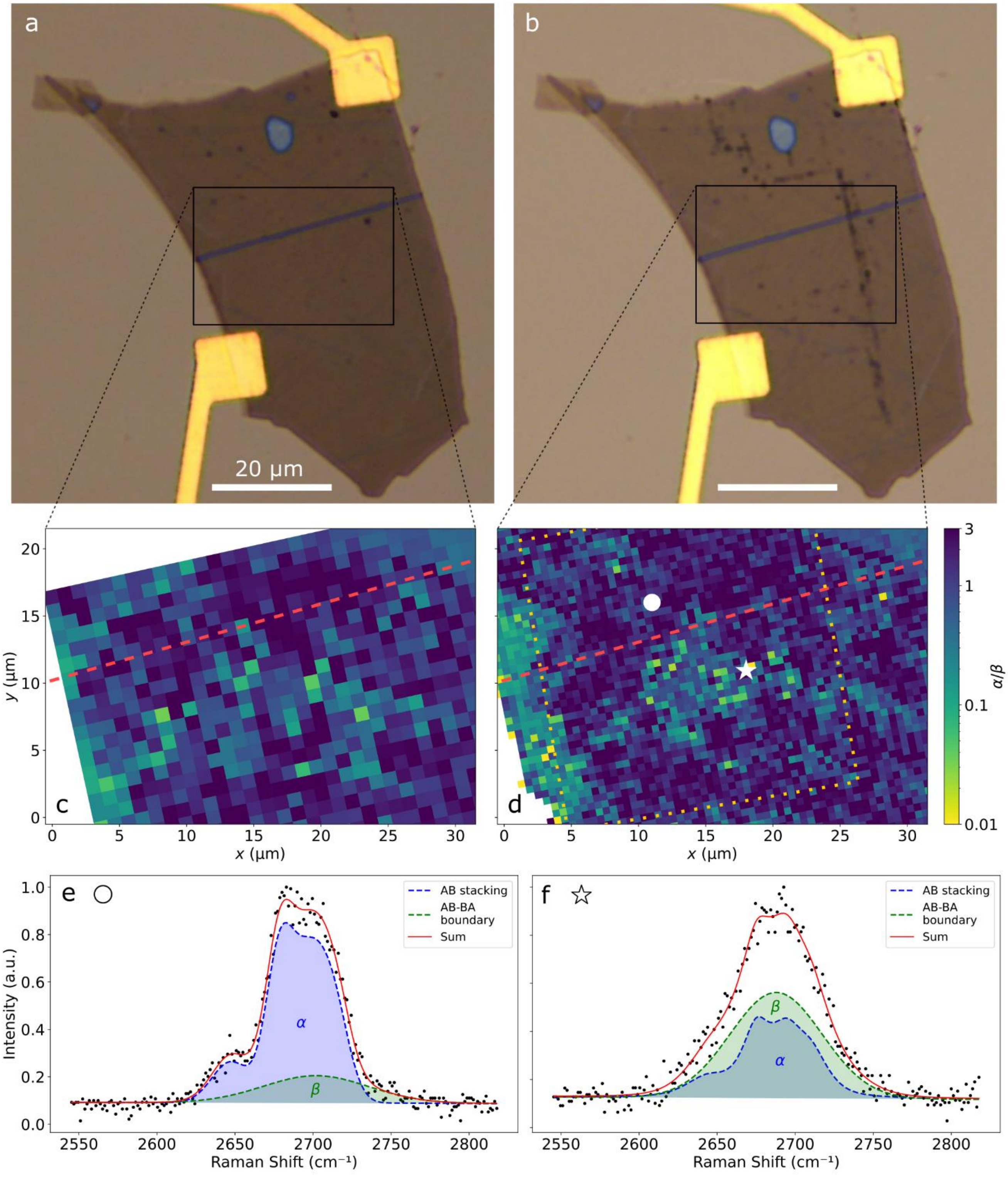

**Figure S1** – **a, b** Optical images of the device before and after AFM-assisted brooming, respectively. The scalebar is the same for both panels. **c, d** Raman-derived maps framed in the same area of the ARPES acquisition displayed in Fig. 3a, the different graphical items having the same meaning. The map is built by computing the α/β ratio between the contributions of the Bernal-stacked areas and the strain soliton regions (AB-BA boundaries) to the 2D peak of TBG. Since the two areas in the moiré superlattice are present in different relative amount depending on the twist angle, the procedure allows to initially identify promising near-magic-angle domains by considering α/β values below 0.1, from geometric consideration. Experimentally, the ratio has been reported to saturate to 0.01 immediately above the magic angle[7], a situation which, however, almost never occurs in our sample. On the other end of the spectrum, values of 1 and above strongly point to Bernal-relaxed domains. The method is a more robust alternative to the simple search for areas maximizing the FWHM of the 2D peak by fitting a single Voigt lineshape to an otherwise complex envelope. **e, f** Two examples of fitting of a strain soliton-associated single Voigt and a Bernal-like contribution (four relatively constrained peaks, calibrated on a Bernal bilayer reference sample employing the same laser and spectrometer) to the 2D-peak spectral range of the Raman map acquired on the broomed sample. The data comes from locations marked by the circle and star in panel d, which are the same as in Fig 3. The α/β ratio yielded by the fitting procedure in each point produces the colormap in panel d, where the assignment to Bernal or near-MA configuration is consistent with ARPES data.

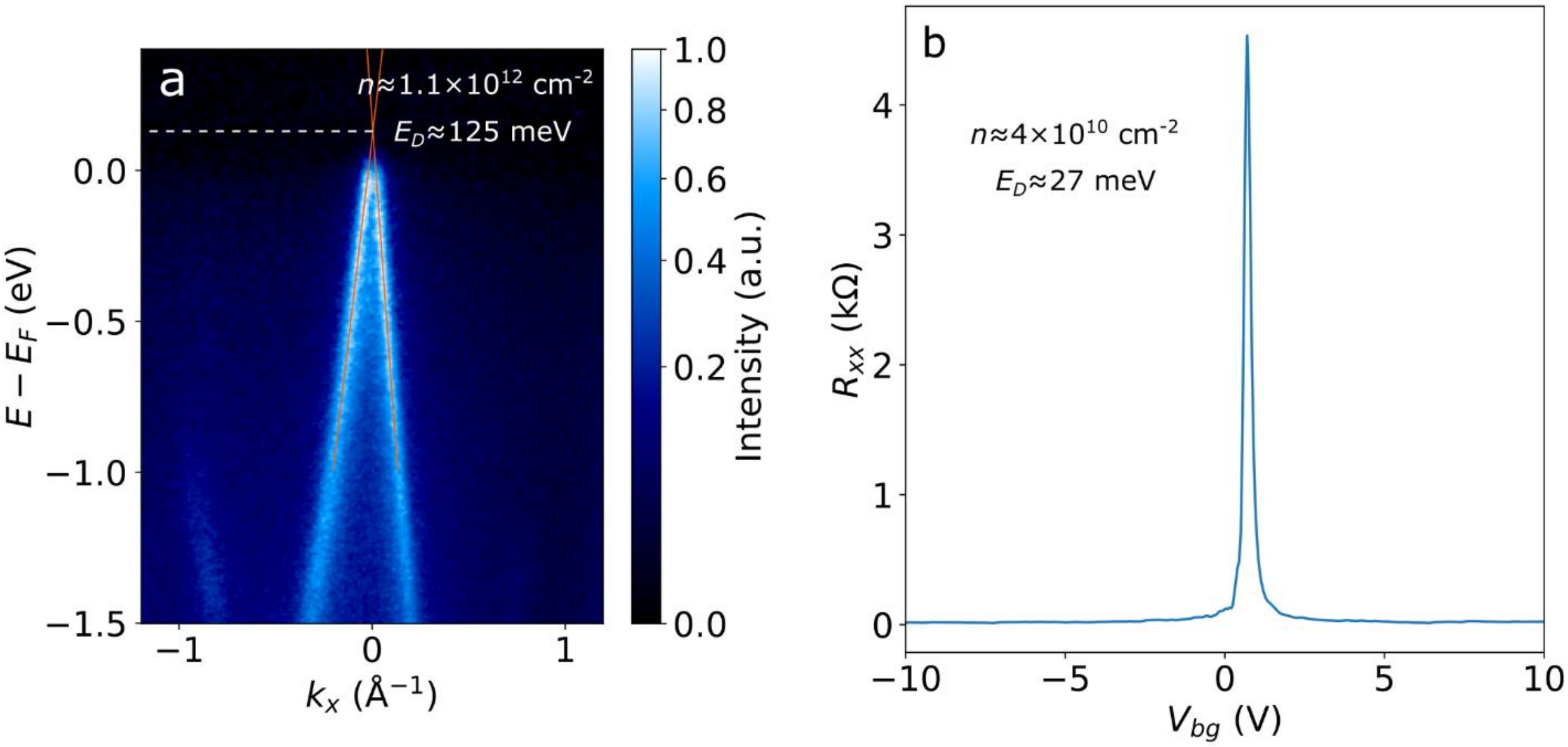

**Figure S2** – **a** ARPES signal acquired in a 30°-TBG region of the device in the main text. The snapshot is taken through the Dirac point of the lower layer (part of the Dirac cone belonging to the upper, twisted layer is visible in the bottom-left corner). The graphene appears p-doped of about 125 meV, corresponding to a carrier density of ~$1.1\times10^{12}$ cm$^{-2}$. This value is influenced by charge redistribution between the two graphene layers. Following Piccinini *et al.*,[13] we consider capacitive coupling between the two graphene layers and induction of charge from the underlying hBN/SiC substrate, obtaining a carrier concentration of ~$0.8\times10^{12}$ cm$^{-2}$ in the upper layer (note that these estimates are an upper bound: if some of the charge is induced from the top surface, then the graphene would acquire less carriers). Therefore, the total carrier concentration is (at most) ~$1.9\times10^{12}$ cm$^{-2}$, which is smaller than the doping required to completely fill with holes the flat band of MATBG, approximately equal to $3\times10^{12}$ cm$^{-2}$. **b** Longitudinal resistance of CVD-grown graphene monolayer on hBN, prepared via the same protocols employed for the device in the main text. The back gate (bg) is the degenerately doped Si underneath the thermal oxide; data were acquired at a temperature of 2 K. The original material is confirmed to be minimally p-doped (carrier density approximately equal to $4\times10^{10}$ cm$^{-2}$, corresponding to a deviation in the Fermi level of about 27 meV from perfect neutrality), hence the attribution of the extensive doping observed in photoemission to a synchrotron-mediated effect. Note that the doping of graphene devices is known not to vary significantly from room temperature (where the ARPES experiment is performed) to cryogenic temperatures (where we perform this reference transport measurement). The estimated mobility at 2 K of this CVD-based single-layer device is ~$4\times10^{5}$ cm$^{2}$ V$^{-1}$ s$^{-1}$, with a n* peak-broadening lower than $10^{10}$ cm$^{-2}$.

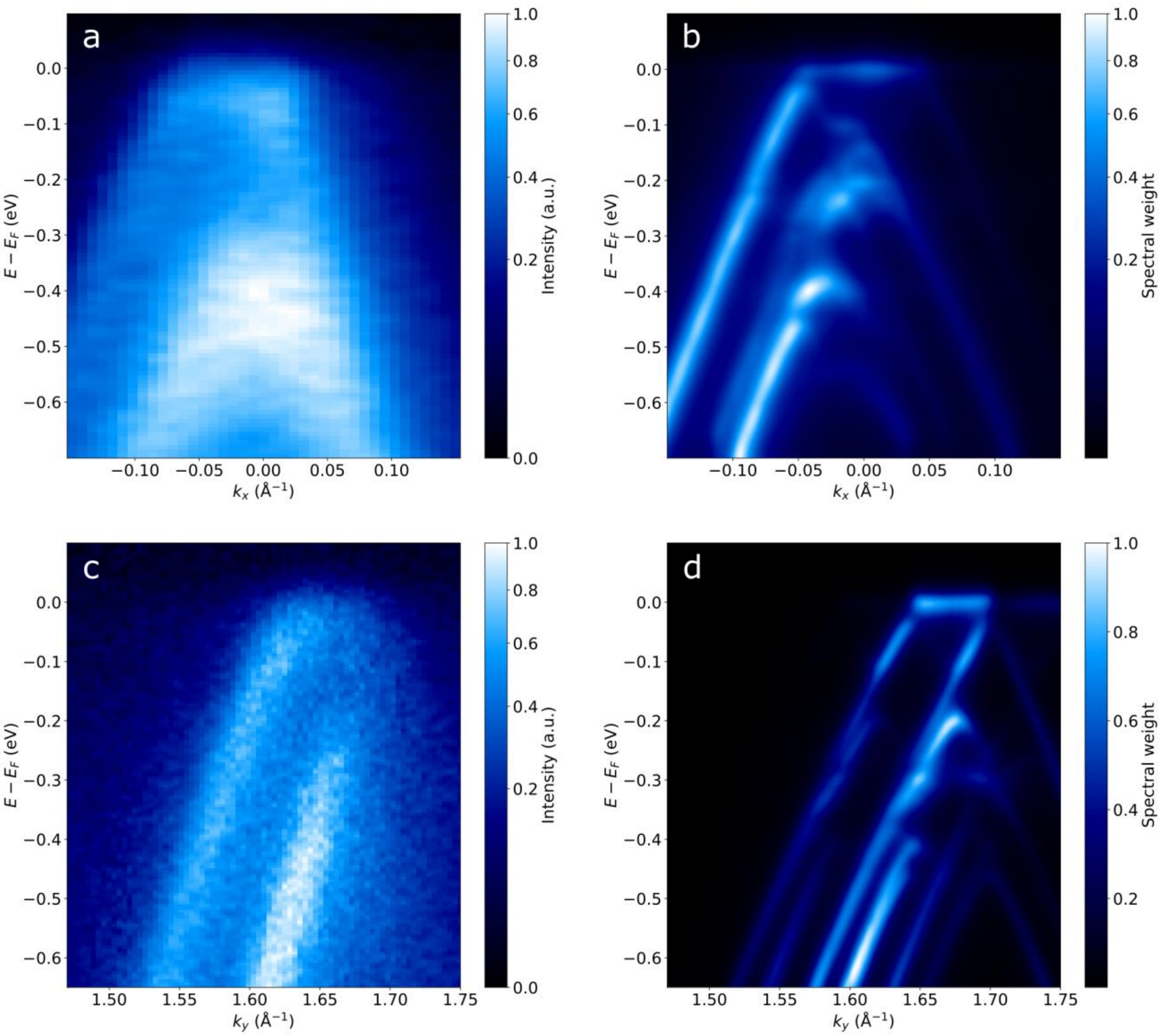

**Figure S3** – Comparison between the band dispersion observed in the sample and the electronic structure of MATBG computed with the Bistritzer-MacDonald model. The main morphological features are shared between the experimental data and the calculation, while differences are due to the thermal-induced noise, the potential inhomogeneities in twist angle within the beam spot size, and the slightly different direction of the cut. **a** nARPES signal acquired within the near-MA domain, along the κmκ' axis, slightly off the Dirac points (same data of Fig. 3b in the main text). **b** Spectral weight computed at 1.09°, along the κmκ' axis, with fixed valley index. **c** nARPES signal acquired within the near-MA domain, along the mγm' axis (same data of Fig. 2b in the main text). **b** Spectral weight computed at 1.09°, along the mγm' axis, with fixed valley index.

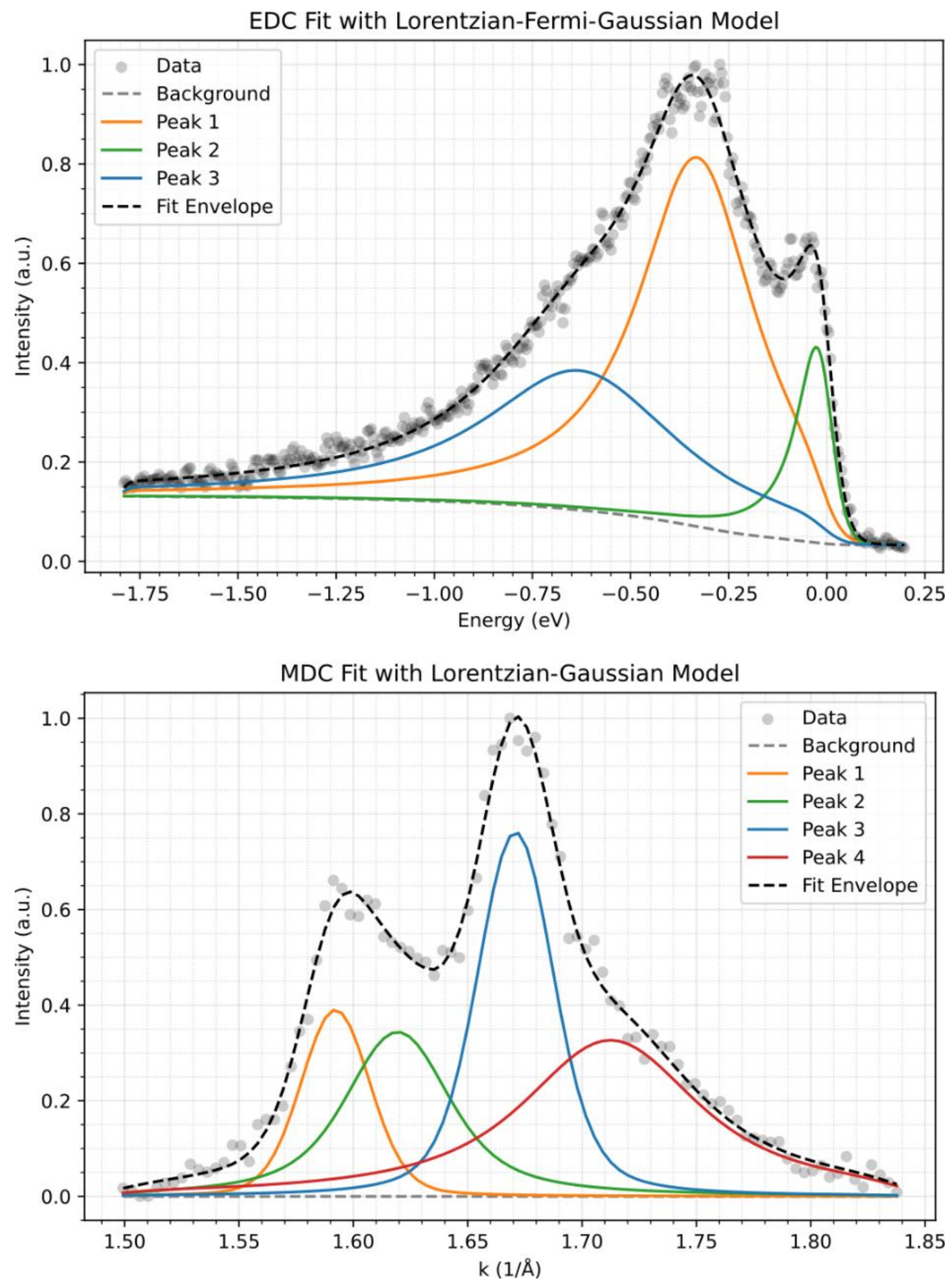

**Figure S4** – (**top**) Fit of a minimally binned EDC extracted at the m point of the moiré mini Brillouin zone (from data in Fig. 2b). The Gaussian broadening of the experimental setup, including the thermal contribution, was calibrated by fitting the Fermi edge of p-doped graphene in Fig. S2a and constrained to the value of 14.69 meV (total energy resolution of ~34 meV). The simplest model allowing to reproduce the acquired signal includes three Voigt lineshapes built as described previously in this document, i.e. as product of a Lorentzian profile and a Fermi distribution function, subsequently convolved with a same-width Gaussian distribution. (**bottom**) Fit of a minimally binned EDC extracted at -0.5 eV from data in Fig. 2b. The model is similar to the previous one, excluding the Fermi-Dirac distribution. A minimum number of four components is necessary. The right-hand side of the spectrum is sensibly less intense and resolvable due to dark corridor effects beyond the K point of single graphene layers.

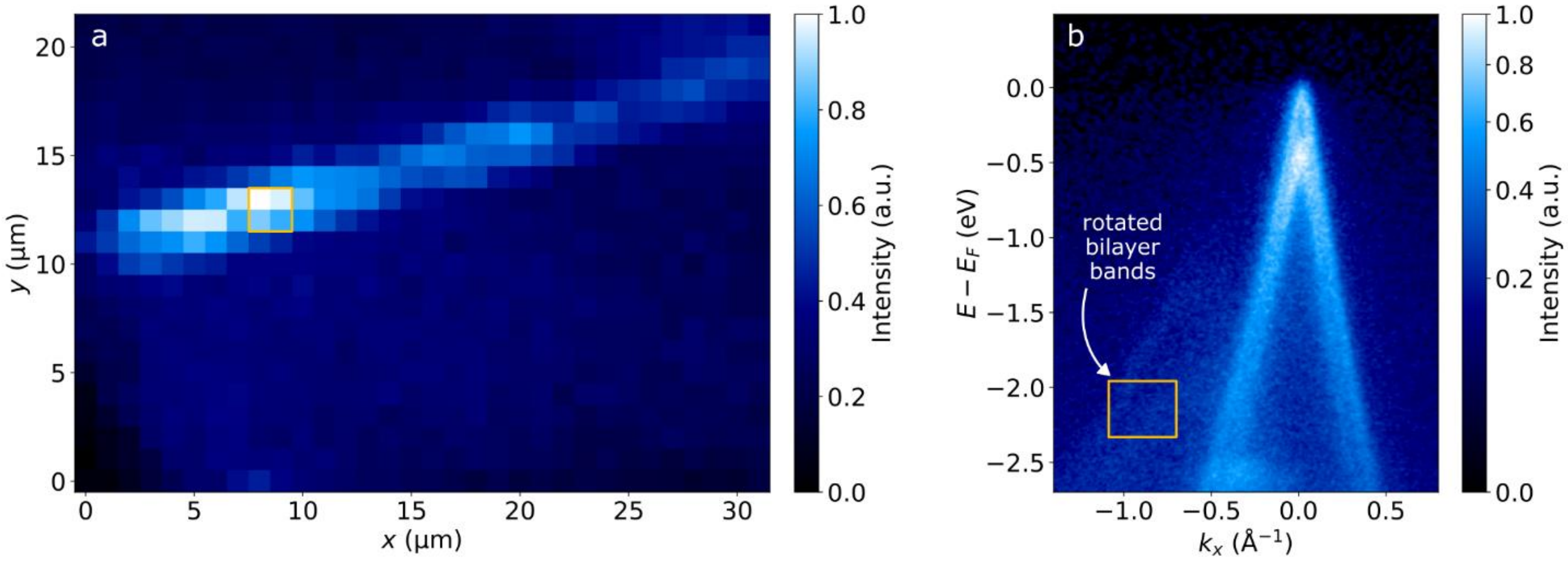

**Figure S5** – Study of the TBG relaxation induced by a wrinkle in the underlying hBN flake. **a** Imaging of the wrinkle obtained *via* the ARPES signal, the field of view is the same as that in Fig. 3a. By integrating the signal from a portion of the momentum and energy axes (orange box in (b)), it is possible to image in real space the structure emitting the signal itself. **b** $k_x$ cut of the band dispersion obtained by integration of the signal in the orange box in (a). To the left of the brighter Bernal bilayer dispersion, it is possible to see the faint signature of another Bernal bilayer dispersion cut off the K point, a circumstance occurring when the whole bilayer orientation changes abruptly on a scale below the resolution limit of the technique and both "domains" overlap their signals in the ARPES snapshot. The discontinuity in the bilayer orientation acts as a strong stress on the crystalline order of both layers, giving rise to deformations that lead to the overall relaxation of the near-MA configuration to AB stacking. Note that the TBG appears to have already relaxed to Bernal stacking even before reaching the precise wrinkle position, as panel (b) was obtained by integration over an area of the sample not yet beyond the wrinkle (box in panel (a)): this suggests that the relaxation of the twist angle is not punctual but rather distributed on a micrometric scale according to a mechanical equilibrium of stress and strain. The high-energy feature below -2.5 eV belongs to the valence band of the misaligned hBN flake.

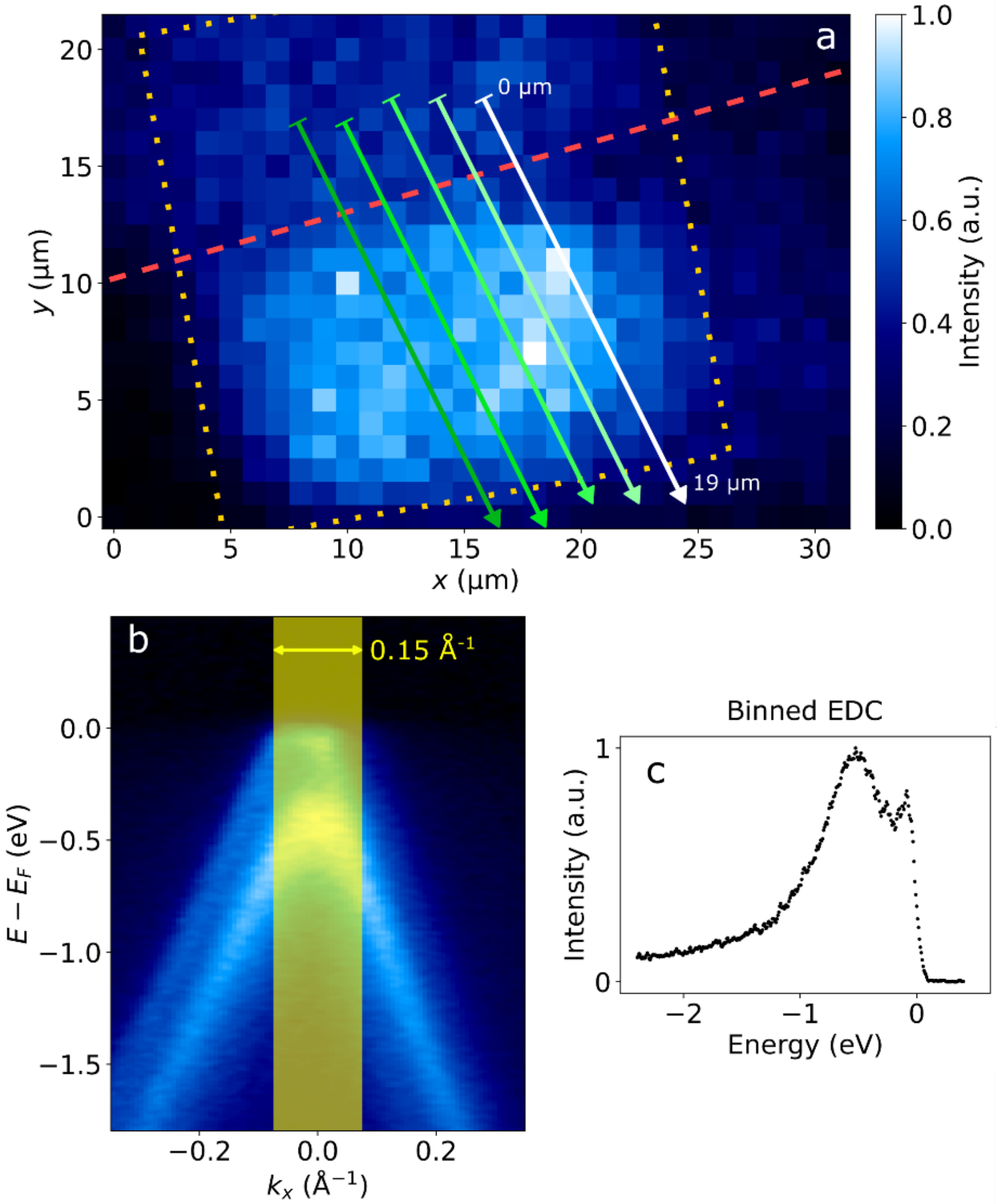

**Figure S6** – Extraction of position-resolved $k_x$-binned EDCs across the near-MA domain. **a** The white arrow marks the direction of extraction of the first series of EDCs, displayed in Fig. 4 in the main text. **b** Employing the data points closest to the chosen axis, EDCs are computed via the binning interval displayed in yellow. **c** Example of a binned EDC, ready for the fitting of its components.

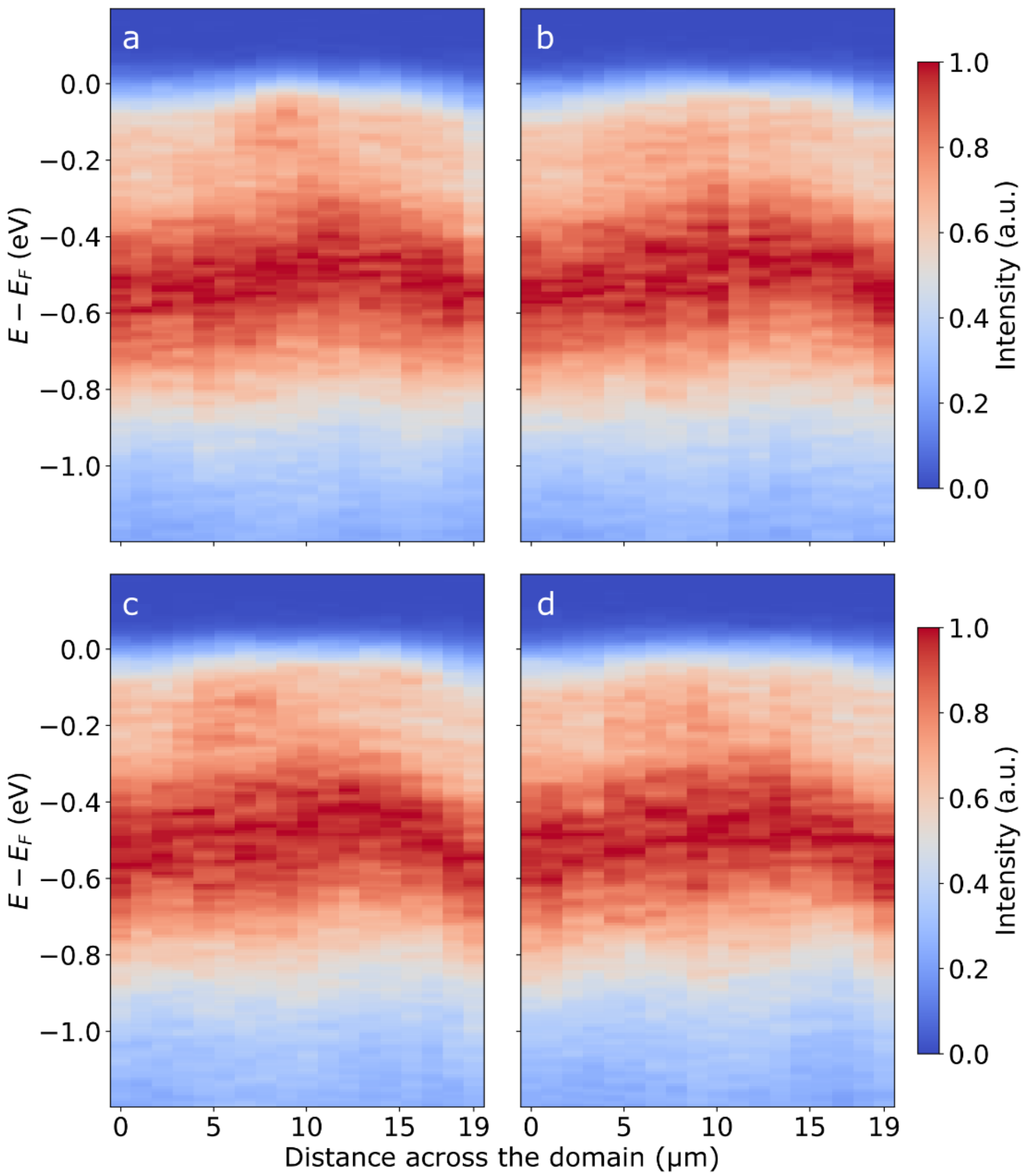

**Figure S7** – The other four binned EDC series extracted across the near-magic-angle domain parallelly to the green arrow in Fig. 3a. The morphology of the signal is analogous to that of the first series reported in Fig. 4a.